# AI-Aided Mapping of the Structure-Composition-Conductivity Relationships of Glass-Ceramic Lithium Thiophosphate Electrolytes

*Haoyue Guo[a], Qian Wang[a], Alexander Urban[a,b,c], and Nongnuch Artrith*[a,b,d]*

[a]Department of Chemical Engineering, Columbia University, New York, NY 10027, USA

[b]Columbia Center for Computational Electrochemistry, Columbia University, New York, NY 10027, USA

[c]Columbia Electrochemical Energy Center, Columbia University, New York, NY 10027, USA

[d]Materials Chemistry and Catalysis, Debye Institute for Nanomaterials Science, Utrecht University, 3584 CG Utrecht, The Netherlands

ABSTRACT Lithium thiophosphates (LPS) with the composition $(Li_2S)_x(P_2S_5)_{1-x}$ are among the most promising prospective electrolyte materials for solid-state batteries (SSBs), owing to their superionic conductivity at room temperature (>$10^{-3}$ S cm$^{-1}$), soft mechanical properties, and low grain boundary resistance. Several glass-ceramic (*gc*) LPS with different compositions and good

Li conductivity have been previously reported, but the relationship between composition, atomic structure, stability, and Li conductivity remains unclear due to the challenges in characterizing non-crystalline phases in experiments or simulations. Here, we mapped the LPS phase diagram by combining first principles and artificial intelligence (AI) methods, integrating density functional theory, artificial neural network potentials, genetic-algorithm sampling, and *ab initio* molecular dynamics simulations. By means of an unsupervised structure-similarity analysis, the glassy/ceramic phases were correlated with the local structural motifs in the known LPS crystal structures, showing that the energetically most favorable Li environment varies with the composition. Based on the discovered trends in the LPS phase diagram, we propose a candidate solid-state electrolyte composition, $(Li_2S)_x(P_2S_5)_{1-x}$ ($x \sim 0.725$), that exhibits high ionic conductivity ($>10^{-2}$ S cm$^{-1}$) in our simulations, thereby demonstrating a general design strategy for amorphous or glassy/ceramic solid electrolytes with enhanced conductivity and stability.

INTRODUCTION

Solid-state-batteries (SSBs) are a prospective alternative to conventional Li-ion batteries (LIBs), in which the flammable liquid electrolytes are replaced with safer solid Li-ion conductors. Additionally, SSBs can potentially enable the use of Li metal anodes and thus significantly higher energy densities.[1–3] Different classes of materials have been investigated as solid electrolytes (SE), including oxides, polymers, phosphates and thiophosphates.[4–7] Among all the prospective SE materials, lithium thiophosphates (LPS) with the composition $(Li_2S)_x(P_2S_5)_{1-x}$ are among the most promising, owing to their superionic conductivity at room temperature ($>10^{-3}$ Scm$^{-1}$), soft mechanical properties, and low grain boundary resistance.[8,9] The implementation of

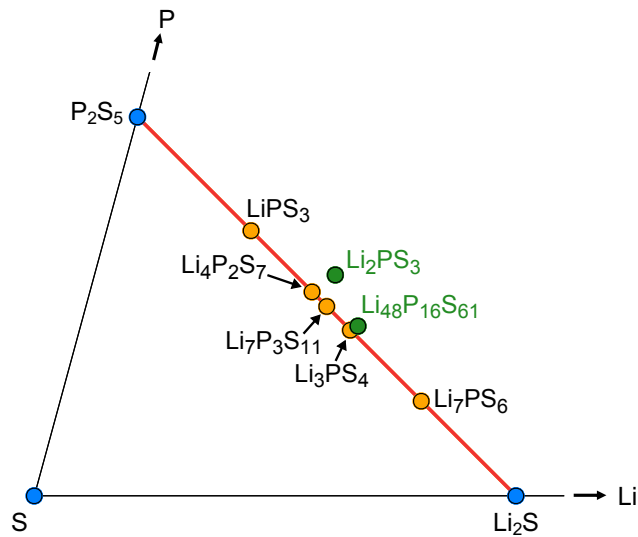

**Figure 1.** Excerpt from the ternary Li–P–S phase diagram showing reported LPS compositions on and near the $Li_2S$–$P_2S_5$ composition line. The materials falling on the right of the red line are sulfur-deficient compositions (green circles).

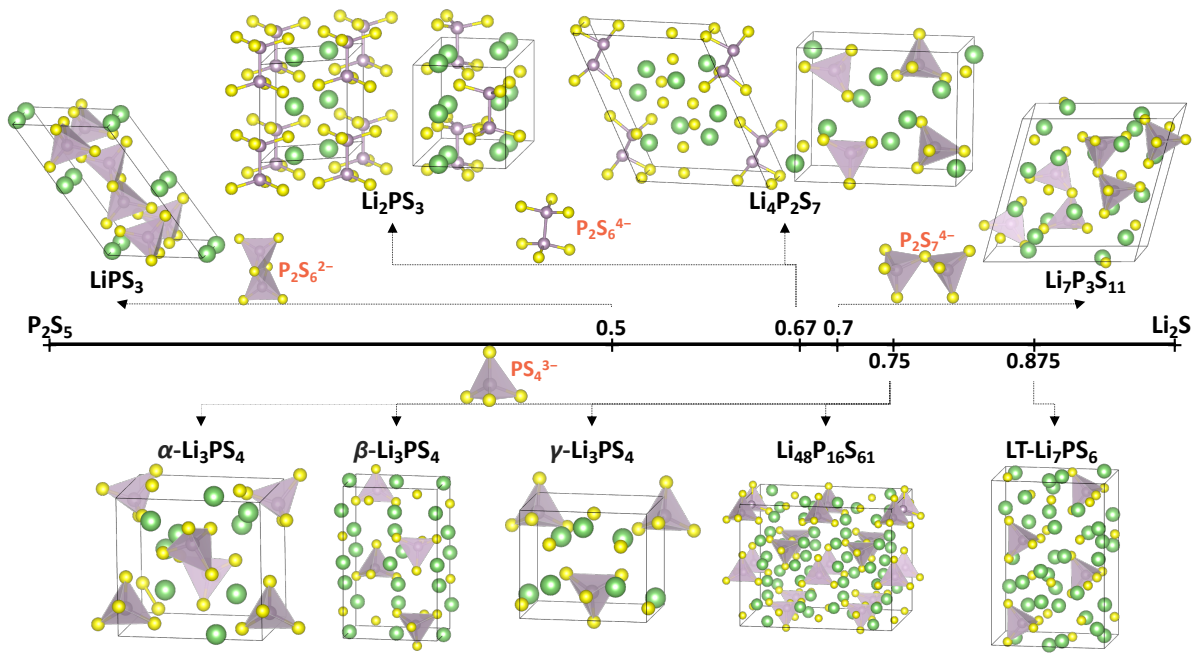

**Figure 2.** Crystal structures of LPS compositions on and near the $Li_2S$–$P_2S_5$ composition line (Li: green; S: yellow; P: purple). The structures are grouped by their local P–S motifs (see **Figure 3**). Note that $Li_2PS_3$ and $Li_{48}P_{16}S_{61}$ do not exactly lie on the $Li_2S$–$P_2S_5$ composition line, as seen in **Figure 1**. Note that the structure of the high-temperature α-$Li_3PS_4$ phase has not been fully resolved in experiment, and our assignment here is speculative.

LPS glasses as SE was first reported in 1980,[10] where it was discovered that the substitution of O with S in phosphates significantly increased the ionic conductivity. In 2006, Mizuno and co-workers observed that the conductivity of LPS materials can be further promoted by partial crystallization of the $Li_2S$–$P_2S_5$ glasses.[11,12] By now, a number of different glass-ceramic (*gc*) LPS compositions have been synthesized and characterized, including $LiPS_3$ (($Li_2S)_{0.5}(P_2S_5)_{0.5}$),[13] $Li_2PS_3$ (($Li_2S)_{0.667}(P_2S_5)_{0.333}$),[14–16] $Li_7P_3S_{11}$ (($Li_2S)_{0.7}(P_2S_5)_{0.3}$),[11,12,17–25] $Li_3PS_4$ (($Li_2S)_{0.7}(P_2S_5)_{0.3}$),[24,26–34] and $Li_7PS_6$ (($Li_2S)_{0.875}(P_2S_5)_{0.125}$),[35] all of which lie on or near the $P_2S_5$–$Li_2S$ composition line in the Li–P–S phase diagram (**Figure 1**).

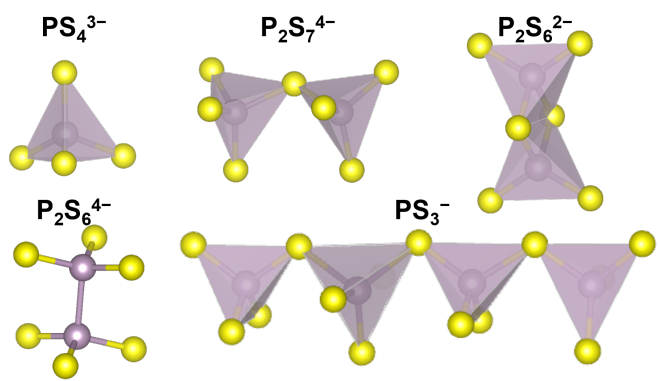

**Figure 3.** P–S anion motifs in different $gc$-LPS: ortho-thiophosphate ($PS_4^{3-}$), pyro-thiophosphate ($P_2S_7^{4-}$), hypo-thiodiphosphate ($P_2S_6^{4-}$), meta-thiodiphosphate ($P_2S_6^{2-}$), meta-thiophosphate ($PS_3^{-}$) (S: yellow; P: purple).

LPS compositions crystallize in several different crystal structures (**Figure 2**) that have been extensively characterized with experimental techniques, such as X-ray powder diffraction (XRD) and nuclear magnetic resonance (NMR)[12,13,15,16,24,35–37] spectroscopy as well as with computational methods.[38–41] Nevertheless, glass-ceramic ($gc$) LPS-based SEs exhibit both crystalline and non-crystalline phases, and the ionic conductivity of such $gc$-LPS materials is significantly influenced by the glassy phases.[41,42] Although the crystal structures and electronic properties of LPS have been thoroughly studied, the relationship between structures and Li conductivity in the $gc$-LPS materials has not been well understood, also due to the limitations of experimental and computational techniques for characterizing non-crystalline phases.

In contrast to crystal structures, glasses lack long-range atomic ordering. It has previously been reported that the energy landscape for ion migration can be impacted by subtle variations in the local structures of LPS,[31,41,43,44] where different local P–S motifs are present depending on the LPS composition. **Figure 3** illustrates the five $P_xS_y^{n-}$ anionic species commonly observed: ortho-thiophosphate ($PS_4^{3-}$), pyro-thiophosphate ($P_2S_7^{4-}$), hypo-thiodiphosphate ($P_2S_6^{4-}$), meta-thiodiphosphate ($P_2S_6^{2-}$), and meta-thiophosphate ($PS_3^-$).[45] Polymeric chains of $PS_3^-$ are only observed in the LPS glasses with low $Li_2S$ contents ($x \leq 0.5$ in **Figure 2**).[45] Glass-ceramics, containing both crystalline and glassy domains, can be synthesized via ball-milling of the crystalline LPS compounds or by nucleating crystallites in glassy materials via heat treatment.[45–47] Although the preparation methods can be dramatically different, the relative ratios of local motifs have been found to be similar as long as the composition remains the same.[45]

Different local P–S motifs can affect the Li sites and therefore change the Li ionic conductivity.[41,43,44] For example, three $Li_3PS_4$ polymorphs, $\alpha$-, $\beta$-, $\gamma$-$Li_3PS_4$, have been synthesized and characterized.[30–32] $\alpha$-$Li_3PS_4$ was formed at high temperatures above 746 K;[31] while $\beta$-$Li_3PS_4$ was first obtained at 573 K[30] and subsequently also at room temperature with other preparation methods.[32] $\gamma$-$Li_3PS_4$ was obtained only at room temperature.[30] Although the local P–S motifs in the three $Li_3PS_4$ polymorphs are exclusively isolated $PS_4^{3-}$ tetrahedra, the phases exhibit different cation arrangements and differ in the orientation of the $PS_4^{3-}$ tetrahedra. Recent theoretical studies proposed that the Li mobility in the $\beta$ phase is increased because of a paddle-wheel mechanism for Li migration that is observed in $\beta$-$Li_3PS_4$ but not in $\gamma$-$Li_3PS_4$.[41,43,44]

Previous computational studies mainly focused on the crystalline LPS phases, such as $Li_2PS_3$,[48–50] $Li_7P_3S_{11}$,[21,51–56] $Li_3PS_4$,[41,43,44,57] and $Li_7PS_6$.[58] The impact of local structure motifs on ionic conductivity in $gc$-LPS has recently been investigated by Sadowski and Albe,[59] who report that structural units do not significantly affect the Li conductivity of the glassy phases. However, the Li migration mechanism remains controversial in the literature, since crystalline $Li_7P_3S_{11}$ exhibits the highest ionic conductivity despite exhibiting corner-shared $PS_4^{3-}$ tetrahedra as local P–S motifs.[21,51–55] In an earlier kinetic study combining reverse Monte Carlo (RMC) modeling and neutron diffraction, it was proposed that the corner-sharing $P_2S_7^{4-}$ shields the positive charge of P due to electron transfer between P and bridging S, therefore suppressing Li conduction.[60–62] However, a later *ab initio* molecular dynamics (AIMD) study found that the flexibility of $P_2S_7^{4-}$ ditetrahedra facilitates Li$^+$ diffusion.[21]

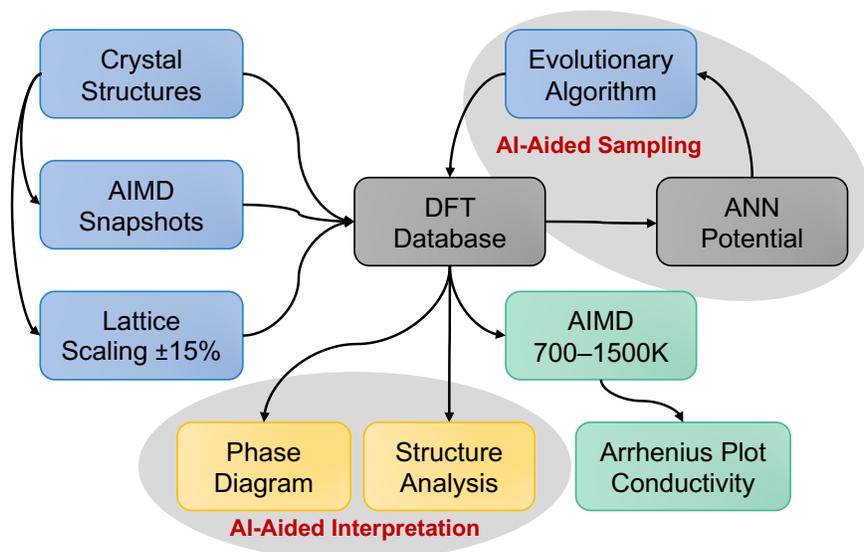

**Figure 4.** Workflow that was used for the AI-aided mapping of the glass-ceramic (*gc*)-LPS phase diagram by combining density-functional theory (DFT) calculations and accelerated sampling with artificial neural network (ANN) potentials and an evolutionary (genetic) algorithm. All final reported results were obtained from either static DFT calculations (yellow boxes) or DFT-based *ab-initio* molecular dynamics (AIMD) simulations (green boxes).

In essence, only few theoretical studies of amorphous/glassy LPS structures have been reported, and the effect of amorphization on Li conduction has not yet been well understood. Conventional density functional theory (DFT) based AIMD simulations alone are limited to relatively small structure models with ~200 atoms, which makes it challenging to investigate amorphous phases without long-range ordering. In some studies, amorphous structures were approximated with moderately sized defect structures or AIMD at high temperatures,[54,57,63–67] which already required significant computational resources. On the other hand, machine learning potentials trained on first-principles reference data can be efficient and accurate for describing amorphous phases with reasonable computation cost.[68–70]

To determine the local atomic structures of *gc*-LPS with varying composition, we mapped the *gc*-LPS phase diagram by integrating DFT,[71] artificial neural network (ANN) potentials,[72] evolutionary/genetic-algorithm (GA) sampling, and AIMD simulations as illustrated by the workflow diagram in **Figure 4**. By varying the compositions along the $Li_2S$–$P_2S_5$ composition line using an (artificial intelligence) AI-aided sampling approach, the phase diagram of *gc*-LPS was completed. For each LPS composition, GA global structure optimizations with an ANN potential were performed to determine low-energy atomic configurations. The relevant near-ground-state structures determined by this sampling approach were recomputed with DFT, and all reported final results are based on DFT. The thermodynamic stability and ionic conductivity of glassy/ceramic phases was correlated with local structural motifs, which allowed identifying structure-composition-conductivity relationships. With machine learning accelerated sampling and AIMD simulations, a candidate solid-state electrolyte composition, $(Li_2S)_x(P_2S_5)_{1-x}$ (x = 0.724), with high ionic conductivity (>$10^{-2}$ S cm$^{-1}$) was identified, which points towards a design strategy for LPS-based SE materials with enhanced conductivity and stability.

## METHODS

*Density Functional Theory Calculations*

All DFT calculations were carried out with the projector-augmented-wave (PAW) method[73,74] and the Perdew–Burke–Ernzerhof (PBE) exchange-correlation functional[75] as implemented in the Vienna *Ab Initio* Simulation Package (VASP)[71,73] and an energy cutoff of 520 eV. Gaussian smearing with a width of 0.05 eV was used, total energies were generally converged better than $10^{-5}$ eV/atom, and the final force on each atom was less than 0.02 eV/Å. The first Brillouin zone was sampled using VASP's fully automatic k-point scheme with length parameter $R_k = 25$ Å.

Amorphous structure models were generated with AIMD simulations of supercells containing 80–128 atoms. In AIMD simulations, a Gamma k-point scheme was employed to reduce the computational cost. The time step for the integration of the equations of motion was set to 1 fs, and the temperature of the system was set to 1200 K using a Nosé-Hoover thermostat.[76] To obtain near-ground-state structures as reference for the machine-learning potential (see below), 150 evenly spaced snapshots were extracted from the AIMD trajectories that were reoptimized with DFT at zero Kelvin via geometry optimizations as described above.

To determine ionic conductivities, ~300 picosecond long AIMD simulations were performed for select compositions (detailed in the results section) after at least 50 picosecond equilibration at the temperatures 700 K, 900 K, 1200 K and 1500 K. The ionic conductivities at room temperature and the activation energies were obtained from Arrhenius extrapolation.[69]

*Machine-Learning Potentials*

All machine-learning potential (MLP) simulations were performed with artificial-neural network (ANN) potentials[77,78] as implemented in the atomic energy network package (ænet).[72,78,79] ANN potentials represent the total energy $E_{\text{tot}}$ of an atomic structure as the sum of atomic energies, $E_{\text{tot}} = \sum_{i}^{N_{\text{atom}}} E_i$, where the atomic energies $E_i$ are predicted by ANNs for a given local atomic environment, and $N_{\text{atom}}$ is the number of atoms in the structure. To be suitable as ANN inputs, local atomic environments, including atomic positions and species, need to be *featurized*, i.e., transformed to a vector representation with constant dimension.[80] In the present work, a Chebyshev basis set with a cutoff of 6.0 Å for the radial expansion (expansion order 18) and a cutoff of 3.0 Å for the angular expansion (expansion order 4) was employed for the featurization of local atomic environments.[78] An ANN architecture with two hidden layers of 15 nodes each and hyperbolic tangent activation functions was employed. The Broyden-Fletcher-Goldfarb-Shanno (BFGS) method[81] was employed for the weight optimization. 10% of the reference data were randomly selected as independent validation set for cross-validation and were not used during training. The training was repeated ten times for 500 training iterations using different randomly initialized weight parameters, and the ANN potential with the lowest validation-set error was selected.

*Genetic Algorithm Sampling*

For accelerated sampling, a specialized ANN potential was trained on a data set containing ~6,000 atomic structures derived from crystalline LPS with lattice parameters scaled by ±15% and perturbed atomic positions from AIMD simulations as described above. The ANN potential yields a root-mean-squared error of 1.4 meV/atom and a mean absolute error of 0.6 meV/atom relative to the DFT reference energies on the validation set. As previously demonstrated for amorphous LiSi alloys and LiPON solid electrolytes,[68,69] specialized ANN potentials constructed based on moderately sized reference data sets can be used in conjunction with DFT for accelerated sampling of amorphous phases.

With the specialized ANN potential, the amorphous phases along the $Li_2S$–$P_2S_5$ composition line were sampled with a genetic-algorithm (GA) as implemented in the atomistic evolution (ævo) package (http://ga.atomistic.net),[68] following previously reported strategies.[68,69] Although glassy phases lack long-range ordering, it can be expected that the local atomic motifs in *gc*-LPS phases resemble those of the known LPS crystalline phases (**Figure 3**). The phase diagram of LPS compositions was therefore constructed by varying the stoichiometry $x$ in $(Li_2S)_x(P_2S_5)_{1-x}$ via removing $Li_2S$ or $P_2S_5$, respectively, from supercells of the known LPS crystal structures. Starting from supercells of the ideal crystal structures of $LiPS_3$, $Li_7P_3S_{11}$, $\beta$-$Li_3PS_4$, $\gamma$-$Li_3PS_4$ and $Li_7PS_6$, either Li and S atoms were removed with a ratio of 2:1, or P and S atoms were removed with a ratio of 2:5, and low-energy configurations were determined with GA sampling. A population size of 32 trials and a mutation rate of 10% were employed. For each composition, at least 10 lowest energy structure models identified with the ANN-GA approach were selected and fully relaxed with DFT to obtain the first principles phase diagram.

*Formation Energy*

For any given structure and composition $(Li_2S)_x(P_2S_5)_{1-x}$ the corresponding formation energy per atom was calculated as

$$E_{\text{f/Atom}} = \frac{E_{(\text{Li}_2\text{S})_x(\text{P}_2\text{S}_5)_{1-x}} - xE_{\text{Li}_2\text{S}} - (1-x)E_{\text{P}_2\text{S}_5}}{7 - 4x}, \quad (1)$$

where $E$ is the total energy of a specific configuration as predicted by DFT; $x$ is the molar fraction of Li$_2$S in the LPS composition; $E_{\text{Li}_2\text{S}}$ and $E_{\text{P}_2\text{S}_5}$ are constant and are equal to the total energy per formula unit of bulk Li$_2$S and P$_2$S$_5$, respectively. For any given composition, the configuration with a lower formation energy is thermodynamically favored at zero Kelvin. The stability of different compositions can be compared by constructing the lower convex hull of the formation energies to obtain the phase diagram.[68]

*Structure similarity and classification*

Low-energy amorphous LPS structures were compared with the known LPS crystal structures by their connectivity of PS$_4$ tetrahedra, following a previous study.[69] In addition, we analyzed structure similarities based on structure *fingerprints*, i.e., each considered structure was transformed to a feature vector with constant dimension. These structure fingerprints were constructed based on the Chebyshev descriptors of local atomic environments, mentioned above in the context of ANN potentials.[78] The local environment of an atom $i$ is represented by a Chebyshev feature vector $\vec{f}_i$. To construct a structure fingerprint $\vec{F}$, the first $K$ moments of the distribution of the atomic feature vectors were calculated, where the $k$-th moment is given by

$$\vec{f}^{(k)} = \frac{1}{N_{\text{atom}}} \sum_i^{N_{\text{atom}}} \left(\vec{f}_i - \langle \vec{f} \rangle\right)^k \quad \text{with } k > 1, \quad (2)$$

and $\langle \vec{f} \rangle = \vec{f}^{(1)}$ is the mean atomic feature vector (the first moment). The structure fingerprint is then the union (i.e., vector concatenation) of the distribution moments, $\vec{F} = \vec{f}^{(1)} \cup \vec{f}^{(2)} \cup ...$, until a maximum moment. In practice, we found that truncating after the second moment already yielded unique structure fingerprints that can distinguish all atomic structures in our database. Atom-type specific structure fingerprints can be constructed by including only atomic feature vectors for the local atomic environments of select atomic species. We made use of this approach by constructing structure fingerprints based on only the local atomic environment of Li atoms. Finally, we reduced the dimension of the structure fingerprints by performing a principal component analysis (PCA) after data standardization, using the PCA and StandardScaler implementations of the *scikit-learn* library.[82] We found 10 principal components to be sufficient, which can explain 85% of the data variance. Hence, each atomic structure in our database could be uniquely represented by a fingerprint vector with 10 components.

Using the structure fingerprints, we define the similarity $S_P$ of two atomic structures as the Pearson correlation coefficient

$$S_P = \frac{\vec{F}_1 \cdot \vec{F}_2}{\|\vec{F}_1\| \, \|\vec{F}_2\|} , \qquad (3)$$

where $\vec{F}_1$ and $\vec{F}_2$ are two (dimension-reduced) structure fingerprints. Furthermore, we performed a cluster analysis of the structure fingerprints using the *k*-means approach as also implemented in *scikit-learn*.[82]

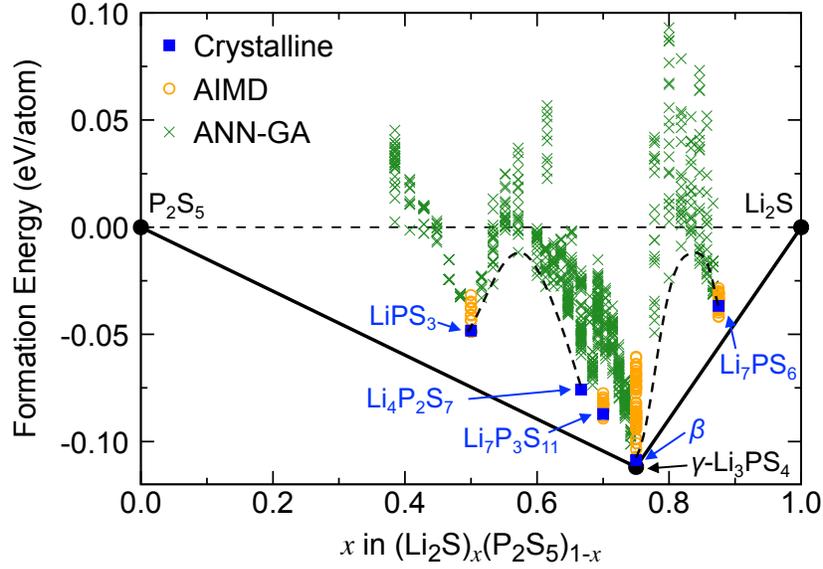

**Figure 5.** Computational LPS phase diagram along the $P_2S_5$–$Li_2S$ composition line. Only the $\gamma$-$Li_3PS_4$ phase lies on the lower convex hull (black solid line) and is thus predicted to be thermodynamically stable at zero Kelvin. Metastable crystalline phases are indicated by blue squares, and structures generated from *ab initio* molecular dynamic (AIMD) simulations and genetic-algorithm (GA) sampling with the ANN potential are shown as orange circles and green crosses, respectively. Two miscibility gaps are indicated with dashed black lines to guide the eye.

RESULTS

*Phase diagram along the $Li_2S$–$P_2S_5$ composition line*

Our computational sampling of the $Li_2S$–$P_2S_5$ composition line started with thirteen LPS crystal structures with the formula units $LiPS_3$,[13] $Li_2PS_3$,[14–16] $Li_4P_2S_7$,[58,64] $Li_7P_3S_{11}$,[17] $\alpha$-$Li_3PS_4$,[31] $\beta$-$Li_3PS_4$,[30,32] $\gamma$-$Li_3PS_4$,[30] $Li_{48}P_{16}S_{61}$[83] and low-temperature (LT)-$Li_7PS_6$[35] that had previously been reported based on experimental characterization and/or theoretical modeling. The crystal structures, which were obtained from the Inorganic Crystal Structure Database (ICSD)[84] and the Materials Project (MP)[85] database, are shown in **Figure 2**. The DFT formation energies of the crystalline LPS phases relative to $Li_2S$ and $P_2S_5$, the endpoints of the composition line, are shown in **Figure 5**. As seen in this phase diagram, only one crystal structure ($\gamma$-$Li_3PS_4$) appears on the lower convex hull of the formation energies and is thus predicted to be thermodynamically stable at zero Kelvin. The previously reported superionic conductors, $\beta$-$Li_3PS_4$[30,32] and $Li_7P_3S_{11}$,[17] are 3.2 meV/atom and 17.2 meV/atom above the convex hull, indicating that they are metastable at zero Kelvin. However, the energy difference between $\beta$-$Li_3PS_4$ and $\gamma$-$Li_3PS_4$ is small (3.2 meV/atom) compared to the thermal energy per degree of freedom at room temperature (~26 meV), so that it is plausible that the $\beta$ polymorph can be thermodynamically stable at room temperature. Note that the crystal structure of $Li_4P_2S_7$[58,64] is a theoretical prediction from the literature and has not been characterized experimentally yet, which is consistent with its comparatively high decomposition energy of 23.5 meV/atom in our phase diagram.

Also shown in the phase diagram of **Figure 5** are structures that were generated using the ANN-GA sampling methodology described in the methods section by removing $Li_2S$ or $P_2S_5$ from supercells of the crystal structures. This composition sampling yielded low-energy structures with structural disorder and no symmetry, as one would expect for *amorphous* or *glassy* phases, while still exhibiting local similarities with the parent crystal structures that they were derived from. At zero Kelvin, these glass-ceramic structures are also predicted to be thermodynamically unstable, though they might be stabilized at synthesis temperatures due to their high entropy (entropy control) or via kinetic trapping.

As seen in the phase diagram, the ANN-GA sampling identified two miscibility gaps between $LiPS_3$ and $Li_4P_2S_7$ and between $Li_3PS_4$ and $Li_7PS_6$, respectively. This means, compositions $(Li_2S)_x(P_2S_5)_{1-x}$ with $0.5 < x < 0.667$ and $0.75 < x < 0.875$ will likely phase separate instead of forming a solid solution, in agreement with previous experimental observations (see also the discussion section).[24] However, between $Li_4P_2S_7$ and $Li_3PS_4$ amorphous structures with low energy above the convex hull (< 90 meV/atom) were found. It can therefore be expected that compositions with $0.667 < x < 0.75$ can be more readily synthesized.

*Structural motifs of the sampled LPS phases*

The LPS crystal structures shown in **Figure 2** are composed of a variety of local motifs (**Figure 3**), which have previously been found to affect the ionic conductivity and the Li transport mechanisms.[45] Isolated $PS_4^{3-}$ tetrahedra are mostly observed in the *gc*-LPS compositions with high $Li_2S$ content ($x \geq 0.75$), such as $\alpha$-$Li_3PS_4$, $\beta$-$Li_3PS_4$,[30,32] $\gamma$-$Li_3PS_4$[30] and $Li_7PS_6$.[35] The $P_2S_7^{4-}$ motif, consisting of two corner-sharing $PS_4$ tetrahedra, is the main building block of the $Li_7P_3S_{11}$ crystal structure[17] as well as glassy LPS compositions with x < 0.75. The $P_2S_6^{2-}$ motif, formed by two edge-sharing $PS_4$ tetrahedra, is observed in *gc*-LPS with $x \leq 0.6$ and is the only local motif in $LiPS_3$ crystals.[13] The $P_2S_6^{4-}$ with direct P–P bonding is typically present in *gc*-LPS with $0.6 \leq x \leq 0.7$.[24] Note that the oxidation state of P is +4 only in the $P_2S_6^{4-}$ motif, while it is +5 in all other local motifs. The $P_2S_6^{4-}$ motif also occurs in $Li_2PS_3$,[14–16] which is a sulfur-deficient composition that is not on the $Li_2S$–$P_2S_5$ composition line.

To better understand the local structures of the ANN-GA sampled *gc*-LPS phases, we computed the radial pair distribution functions (RDFs) for P–S and Li–S in *gc*-LPS compositions with $0.385 \leq x \leq 0.867$ as shown in **Figure 6** and **Figure S1**. As seen in the figures, and as expected, the RDFs of the generated *gc*-LPS structures exhibit features of the crystal structure RDFs but show broadened peaks with shifted peak positions. In general, with decreasing amount of Li$_2$S in *gc*-LPS, the main Li–S peak shifts to greater distances, which is caused by the formation of corner-sharing motifs, in agreement with previous reports.[42,64,65] Note that for a large fraction of the *gc*-LPS structures (~1/3) the shape of the RDF differs significantly from the parent structure, and structures with the same composition that were derived from two different parent structures exhibit similar peaks. This is especially evident in the S–S RDF shown in **Figure S1** and indicates that *gc*-LPS with compositions in between the crystalline phases may exhibit multiple different local structural motifs found in the neighboring (by composition) crystalline LPS.

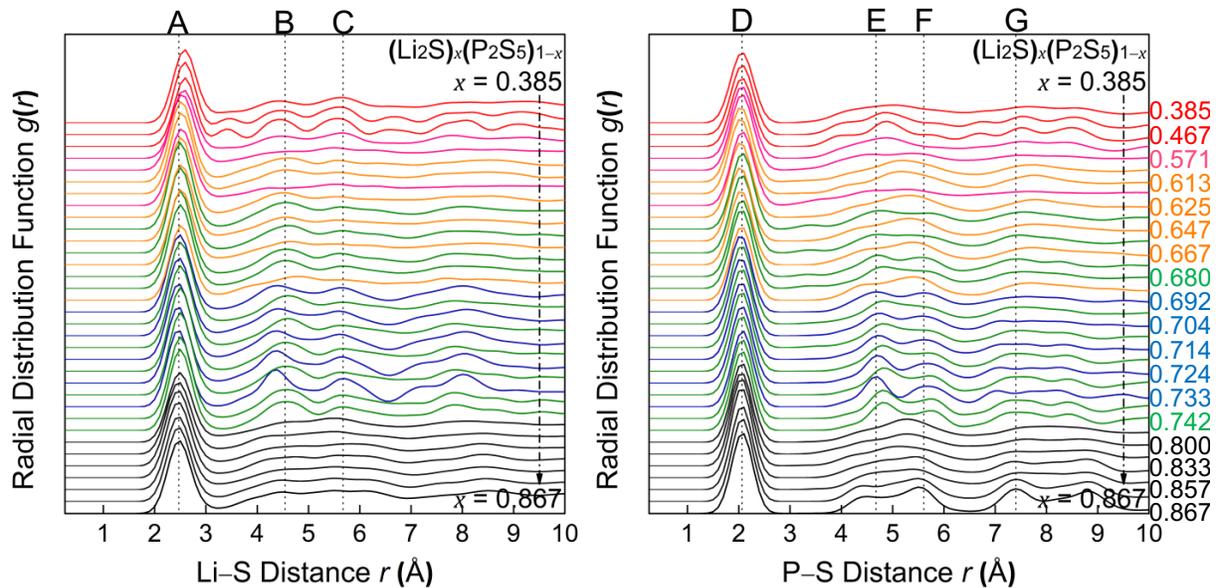

**Figure 6.** Calculated Li–S (left) and P–S (right) radial distribution functions (RDF) of glass-ceramic (*gc*) (Li$_2$S)$_x$(P$_2$S$_5$)$_{1-x}$ (*gc*-LPS) phases with varying composition from x = 0.385 to x = 0.867 (the composition of every other line is labeled on the right). Each line is an average RDF of the ten lowest-energy structures at a specific composition. The *gc*-LPS structures were generated by genetic-algorithm modification of a parent structure (see Methods Section), and the color represents the parent crystal structure (*i.e.*, black: Li$_7$PS$_6$, blue: $\gamma$-Li$_3$PS$_4$, green: $\beta$-Li$_3$PS$_4$, orange: Li$_7$P$_3$S$_{11}$, pink and red: LiPS$_3$). The dashed lines indicate measured RDFs from experiments: Peak A[41,42,62], B[24,41,42], C[24,41], D[15,24,41,62], E[15,24,41], F[24,41], G[24].

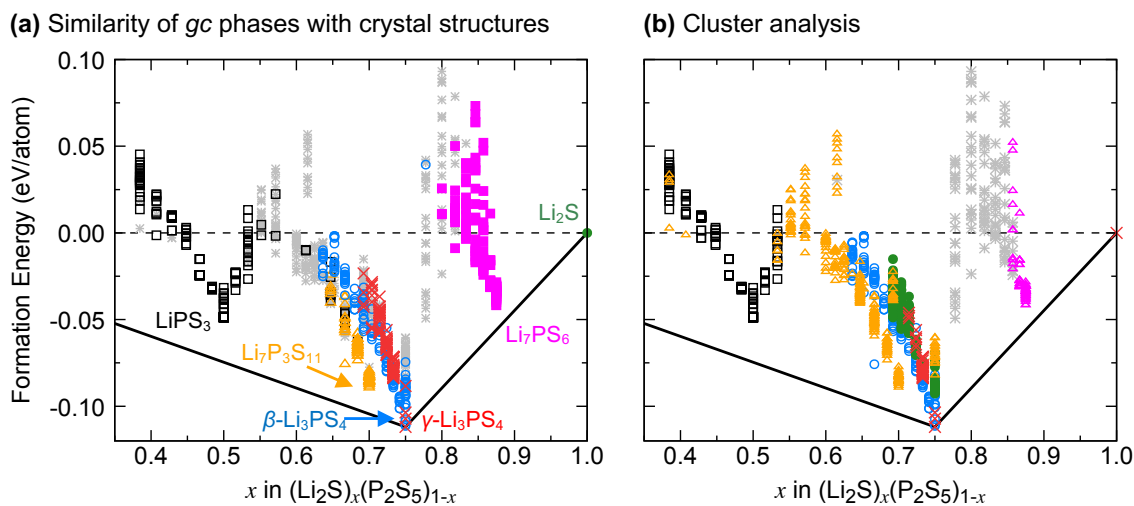

**Figure 7.** Analysis of the local atomic Li environment in the simulated glass-ceramic (*gc*) phases. **(a)** The symbols and color coding indicate the crystal structure that is most similar based on the Pearson correlation of the structural fingerprints. Structures that are not strongly correlated with any crystal structure are shown as gray stars. **(b)** Grouping of similar structures with *k*-means clustering of the Li local atomic environments. The structures within the same cluster are shown with the same symbol and color.

As discussed in the introduction section, the P–S structural building blocks alone cannot explain all differences in Li conductivity, and RDFs capture only one specific structural feature, namely radial correlations. The structural fingerprints introduced in the methods section are more general. **Figure 7.** shows an analysis of the structural fingerprints of all structures in our database to identify and visualize similarities more directly. For this comparison, each structure was represented by a structure fingerprint based on the local atomic environments of all Li atoms, which can be assumed to be an important criterion for Li conductivity.

In **Figure 7. a**, the similarities of each structure with the reference crystal structures $LiPS_3$, $Li_7P_3S_{11}$, $\beta$-$Li_3PS_4$, $\gamma$-$Li_3PS_4$, $Li_7PS_6$, and $Li_2S$ are shown. The Pearson correlation $S_P$ of the structure descriptors (see methods section) was used as a measure of similarity, and structures with $S_P < 0.3$ for all of the crystal structures were considered not to be similar to any of the reference structures. As seen in the figure, structures derived from either $LiPS_3$ or $Li_7PS_6$ remain similar to their parent structure during sampling. However, trends are more complicated for compositions near $Li_3PS_4$ ($x = 0.75$). Within the narrow composition range $0.70 \leq x \leq 0.75$, the structures closest to the ground-state hull change in character from $Li_7P_3S_{11}$ to structures that are similar to $\beta$-$Li_3PS_4$ to $\gamma$-$Li_3PS_4$.

Instead of classifying the sampled glass-ceramic structures by their similarity to reference crystal structures, **Figure 7. b** shows the result of an unsupervised classification of Li environments using $k$-means clustering. The predicted grouping resembles the one shown in **Figure 7. a** but with clearer trends in phase stabilities. At the composition $Li_3PS_4$, the cluster analysis finds that the Li environment changes with increasing energy, which we can attribute to the $\gamma$, $\beta$, and $\alpha$ polymorphs. At high energies above the ground state hull, a fourth class of Li environment is found of which $Li_7P_3S_{11}$ is also a member, though it is unlikely that these structures can be synthesized at any conditions.

*Li conductivity*

The cluster analysis of Li atom environments discussed in the previous section indicates that *gc*-LPS with compositions between $Li_7P_3S_{11}$ ($x = 0.70$) and $Li_3PS_4$ ($x = 0.75$) should exhibit the same type of Li environments as the superionic conductor $\beta$-$Li_3PS_4$. To determine if this similarity also translates to Li conductivity, we performed AIMD simulations for a glass-ceramic LPS with composition $gc$-$Li_{42}P_{16}S_{61}$ ($x = 0.724$), the two neighboring crystalline phases ($\beta$-$Li_3PS_4$ and $Li_7P_3S_{11}$), and a composition outside the target range, $gc$-$Li_{38}P_{24}S_{79}$ ($x = 0.613$), for comparison. The ionic conductivities at room temperature were obtained from Arrhenius extrapolation (**Figure 8a** and **Figure S2**) and are compiled in **Table 1**. The table also shows measured ionic conductivities in $gc$-LPS from the literature, which are sensitive with respect to the experimental conditions, $e.g.$, temperature and pressure. Samples prepared under different conditions may exhibit different local motifs, leading to a wide range of measured conductivities.[21–23]

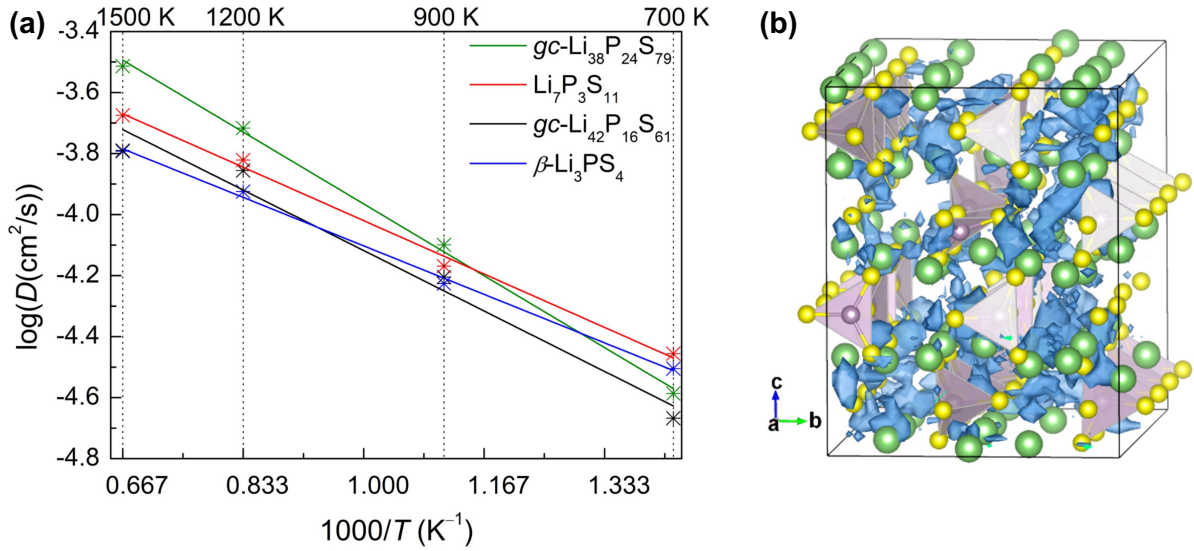

**Figure 8. (a)** Arrhenius plot of the calculated diffusivities from AIMD simulations at elevated temperatures (700 K, 900 K, 1200 K and 1500 K) of selected $gc$-LPS compositions ($gc$-$Li_{38}P_{24}S_{79}$, $gc$-$Li_{42}P_{16}S_{61}$, $Li_7P_3S_{11}$ and $\beta$-$Li_3PS_4$). **(b)** Isosurface of the probability density distribution (blue) $P(\mathbf{r})$ of $Li^+$ ions in $gc$-$Li_{42}P_{16}S_{61}$ at 700 K (Li: green; S: yellow; P: purple).

As shown in **Table 1**, our predicted ionic conductivity and activation energy in crystalline Li$_7$P$_3$S$_{11}$ is in good agreement with previously reported experimental measurements and theoretical calculations. The differences are greater for the $\beta$-Li$_3$PS$_4$ phase, where the agreement with previous simulations is good but predicted conductivities are significantly greater than those observed in experiments. This has to be expected, since the metastable $\beta$ phase is more challenging to characterize experimentally as well as in simulations. Hence, the data for the $\beta$ phase is subject to greater uncertainties.

The ionic conductivity of $gc$-Li$_{42}$P$_{16}$S$_{61}$ is high (33.1 mS cm$^{-1}$) and lies between the conductivities of crystalline Li$_7$P$_3$S$_{11}$, 46.9 mS cm$^{-1}$, and $\beta$-Li$_3$PS$_4$, 14.3 mS cm$^{-1}$. In comparison, the other amorphous phase, $gc$-Li$_{38}$P$_{24}$S$_{79}$ ($x = 0.613$) has a significantly lower ionic conductivity of 3.45 mS cm$^{-1}$ and higher activation energy of 0.282 eV (**Table 1**), showing that non-crystallinity alone is not responsible for the high conductivity. Note that energetically $gc$-Li$_{42}$P$_{16}$S$_{61}$ is only 28.0 meV/atom above the ground-state hull and is likely synthesizable, whereas $gc$-Li$_{38}$P$_{24}$S$_{79}$ lies in a miscibility gap (70.5 meV/atom above the hull) in the phase diagram (**Figure 5**) and is highly unstable, so that the composition would likely phase separate on longer time scales.

DISCUSSION

In the present work, we mapped the phase stability and structure of glass-ceramic lithium thiophosphates along the Li$_2$S–P$_2$S$_5$ composition line. Our calculations identified two miscibility gaps in the composition ranges (Li$_2$S)$_x$(P$_2$S$_5$)$_{1-x}$ with $0.5 \leq x \leq 0.667$ and $0.75 \leq x \leq 0.875$, predicting that solid solutions with such compositions would be challenging to synthesize and likely to phase separate at room temperature. Dietrich et al. previously conducted an experimental study of glass-ceramic LPS compounds with $0.6 \leq x \leq 0.8$ and found that LPS($x = 0.8$) phase separates into $Li_3PS_4$ ($x = 0.75$) and $Li_2S$ ($x = 1.0$),[24] in agreement with our prediction. However, the same authors reported the successful preparation and characterization of LPS($x = 0.6$), which should also be unstable based on the calculated phase diagram. A possible explanation for this discrepancy could be sulfur deficiency in the compositions, since $Li_4P_2S_6$ is a known decomposition product of $gc$-$Li_4P_2S_7$[13] and an attractor in the phase diagram (see **Figure 1**). The impact of such off-stoichiometries deserves a more detailed study in the future.

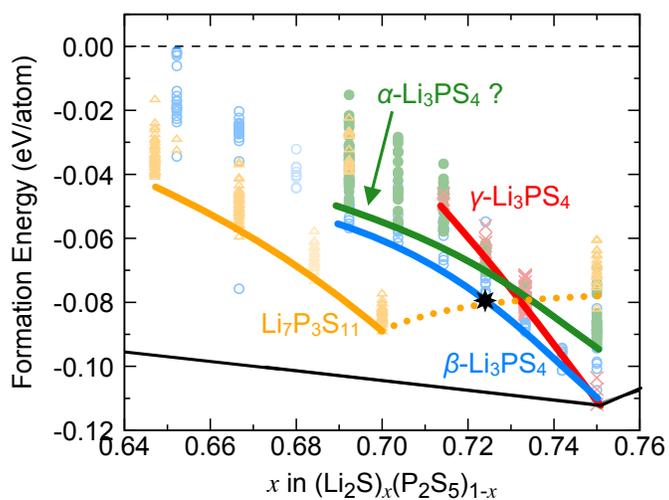

**Figure 9.** Analysis of the LPS phase diagram near the composition $Li_3PS_4 = (Li_2S)_{0.75}(P_2S_5)_{0.25}$, based on the cluster analysis of **Figure 7**. The energetic order of structures with Li environments similar to the $\beta$ and $\gamma$-$Li_3PS_4$ changes as the $Li_2S$ content decreases, and structures that are similar to $\gamma$-$Li_3PS_4$ are destabilized relative to those similar to $\beta$-$Li_3PS_4$. The identified glass-ceramic phase with good Li conductivity, $gc$-$Li_{42}P_{16}S_{61}$, is indicated by a star. Note that the structure of the high-temperature $\alpha$-$Li_3PS_4$ phase has not been fully resolved in experiment, and our assignment here is speculative.

The calculated phase diagram shows that the superionic LPS compounds are metastable and therefore prone to decomposition, which is in agreement with previous experimental and computational work discussed in the introduction section. A particular challenge is that the $\beta$-$Li_3PS_4$ polymorph, a superionic Li conductor, is unstable compared to the $\gamma$-$Li_3PS_4$ polymorph, which exhibits poor Li conductivity. The cluster analysis of Li environments (**Figure 7.** ) points towards an opportunity, since Li environments similar to those in $\beta$-$Li_3PS_4$ become stable compared to those of the $\gamma$ phase when the composition is slightly altered from the ideal $Li_3PS_4$ ($x = 0.75$) to $x < 0.75$. This relative destabilization of the $\gamma$ phase is visualized in **Figure 9**. Indeed, our AIMD simulations confirm that the glass-ceramic *gc*-$Li_{42}P_{16}S_{61}$ ($x = 0.724$) exhibits a high Li conductivity of 33 mS cm$^{-1}$. The RDF analysis of **Figure 6** further shows that the P–S and Li–S distribution in *gc*-$Li_{42}P_{16}S_{61}$ derived from $\beta$-$Li_3PS_4$ still resembles that of the parent phase. As seen in **Figure 8b**, the *gc*-$Li_{42}P_{16}S_{61}$ structure exhibits both well-ordered as well as disordered domains, and the Li probability distribution is greater in the ordered regions. This further indicates that reminiscence of the crystalline phase is important for Li conductivity in this *gc*-LPS composition.

Similar to the known crystalline LPS superionic conductors, the here identified LPS composition is also metastable and thermodynamically unstable with respect to decomposition into $P_2S_5$ and $Li_3PS_4$ at zero Kelvin. However, unlike crystalline phases, the predicted glass-ceramic *gc*-$Li_{42}P_{16}S_{61}$ would benefit from entropy stabilization at finite temperatures. Furthermore, and unlike other glass-ceramic Li conductors, the desired phase with $\beta$-$Li_3PS_4$-like Li environments is predicted to be the lowest in energy at the composition $(Li_2S)_x(P_2S_5)_{1-x}$ with $x = 0.724$, which means that the phase, if it can be synthesized, could be expected to be shelf-stable at room temperature.

Finally, we stress that our computational study is subject to approximations, and an experimental confirmation is warranted. The most significant approximation in the present study is the generation and representation of glass-ceramic phases, which was necessarily limited to comparatively small structure sizes and non-exhaustive sampling. Though, based on previous work,[68,69] ANN-potential accelerated sampling yielded a sufficiently good approximation of the true LPS composition and structure space that the predicted phase diagram and the identified trends in Li environments can be expected to be robust. Another limitation of the present study is that it only considered the $Li_2S$–$P_2S_5$ composition line, even though sulfur-deficient LPS have been reported. The impact of such off-stoichiometries, alluded to in the above discussion, deserves its own investigation.

CONCLUSIONS

We mapped the phase diagram of lithium thiophosphate, $(Li_2S)_x(P_2S_5)_{1-x}$, solid electrolytes using first-principles calculations with AI-aided sampling and structure similarity analysis. The phase diagram exhibits two pronounced miscibility gaps, so that compositions with $0.5 < x < 0.667$ and $0.75 < x < 0.875$ are prone to phase separation at room temperature even if they can be synthesized. We showed that glassy/ceramic phases with compositions $0.70 < x < 0.75$ are more likely to be stable because of their lower decomposition energies and exhibit Li sites with similar local structural environments as those in the superionic conductor $\beta$-$Li_3PS_4$. This led us to propose a candidate solid-state electrolyte composition, $(Li_2S)_x(P_2S_5)_{1-x}$ with $x = 0.724$, that exhibits high ionic conductivity ($>10^{-2}$ S cm$^{-1}$) in simulations, demonstrating a design strategy for glassy or amorphous solid-electrolyte materials with good conductivity and stability.

TABLES

**Table 1.** Comparison of calculated activation energy and Li conductivity of selected *gc*-LPS phases (*i.e.*, *gc*-Li$_{38}$P$_{24}$S$_{79}$, *gc*-Li$_{42}$P$_{16}$S$_{61}$, Li$_7$P$_3$S$_{11}$ and *β*-Li$_3$PS$_4$) with experimental measurements.

| x | Formula | Moiety | Activation Energy (eV) | | | Ionic Cond. RT (mS cm$^{-1}$) | | |
|---|---|---|---|---|---|---|---|---|
| | | | Our AIMD | Ref. AIMD | Exp. | Our AIMD | Ref. AIMD | Exp. |
| 0.613 | *gc*-Li$_{38}$P$_{24}$S$_{79}$ | P$_2$S$_7^{4-}$, PS$_4^{3-}$ | 0.282 | N/A | N/A | 3.45 | N/A | N/A |
| 0.7 | Li$_7$P$_3$S$_{11}$ | P$_2$S$_7^{4-}$, PS$_4^{3-}$ | 0.189 | 0.189[55] 0.187[21] 0.17[54] 0.38[40] | 0.187[11] 0.124[12] 0.145[18] 0.176[20] 0.18~0.209[21] 0.29~0.425[22] 0.289~0.401[23] 0.451[24] | 46.9 | 57[21] 72.16[54] | 3.2[11,12] 4.1[18] 5.2[19] 17[20] 1.3~11.6[21] 0.022~8.6[22] 0.05~4[23] |
| 0.724 | *gc*-Li$_{42}$P$_{16}$S$_{61}$ | PS$_4^{3-}$ | 0.208 | N/A | N/A | 33.1 | N/A | N/A |
| 0.75 | *β*-Li$_3$PS$_4$ | PS$_4^{3-}$ | 0.236 | 0.1, 0.35[65] 0.23[40] 0.35[44] 0.22, 0.25[41] | 0.49[27] 0.352[28] 0.16[31] 0.356[32] 0.399[24] | 14.3 | 4.35[57] 7[41], 19[41] | 0.2[28] 0.16[32] 0.28[24] |

## ASSOCIATED CONTENT

**Supporting Information**. The bulk modulus, RDF and MSD.

The Supporting Information is available free of charge at https://pubs.acs.org/doi/XXX.

*Code availability*

This work made use of the free and open-source atomic energy network (ænet) package. The ænet source code can be obtained either from the ænet website (http://ann.atomistic.net) or from GitHub (https://github.com/atomisticnet/aenet). The evolutionary algorithm package, ævo, can also be obtained from the GitHub (https://github.com/atomisticnet/aevo).

*Data availability*

The reference LPS datasets can be obtained from the Materials Cloud repository (https://doi.org/XXX). The dataset contains atomic structures and interatomic forces in the XCrySDen structure format (XSF),[86] and total energies are included as additional meta information.

## AUTHOR INFORMATION

">
**Corresponding Author**

\* Nongnuch Artrith

E-mail: n.artrith@uu.nl
">
**Funding Sources**

This work was funded by the U.S. Department of Energy (DOE), Office of Energy Efficiency and Renewable Energy, Vehicle Technologies Office, Contract No. DE-SC0012704.

## ACKNOWLEDGMENT

This work was funded by the U.S. Department of Energy (DOE), Office of Energy Efficiency and Renewable Energy (EERE), Vehicle Technologies Office (VTO), Contract No. DE-SC0012704, Advanced Battery Materials Research program (Tien Duong, Program Manager).
">26

The project used resources at the Scientific Data and Computing Center of the Computational Science Initiative at Brookhaven National Laboratory under Contract No. DE-SC0012704. Computations also made use of the Extreme Science and Engineering Discovery Environment (XSEDE), which is supported by National Science Foundation grant number ACI-1053575 (allocation no. DMR14005). We also acknowledge computing resources from Columbia University's Shared Research Computing Facility project, which is supported by NIH Research Facility Improvement Grant 1G20RR030893-01, and associated funds from the New York State Empire State Development, Division of Science Technology and Innovation (NYSTAR) Contract C090171, both awarded April 15, 2010. The authors thank Feng Wang, Deyu Lu, Shinjae Yoo, and Zhoulin Xia for insightful discussions.

# Supporting Information

# AI-Aided Mapping of the Structure-Composition-Conductivity Relationships of Glass-Ceramic Lithium Thiophosphate Electrolytes

*Haoyue Guo[a], Qian Wang[a], Alexander Urban[a,b,c], and Nongnuch Artrith*[a,b,d]*

[a]Department of Chemical Engineering, Columbia University, New York, NY 10027, USA
[b]Columbia Center for Computational Electrochemistry, Columbia University, New York, NY 10027, USA
[c]Columbia Electrochemical Energy Center, Columbia University, New York, NY 10027, USA

[d]Materials Chemistry and Catalysis, Debye Institute for Nanomaterials Science, Utrecht University, 3584 CG Utrecht, The Netherlands

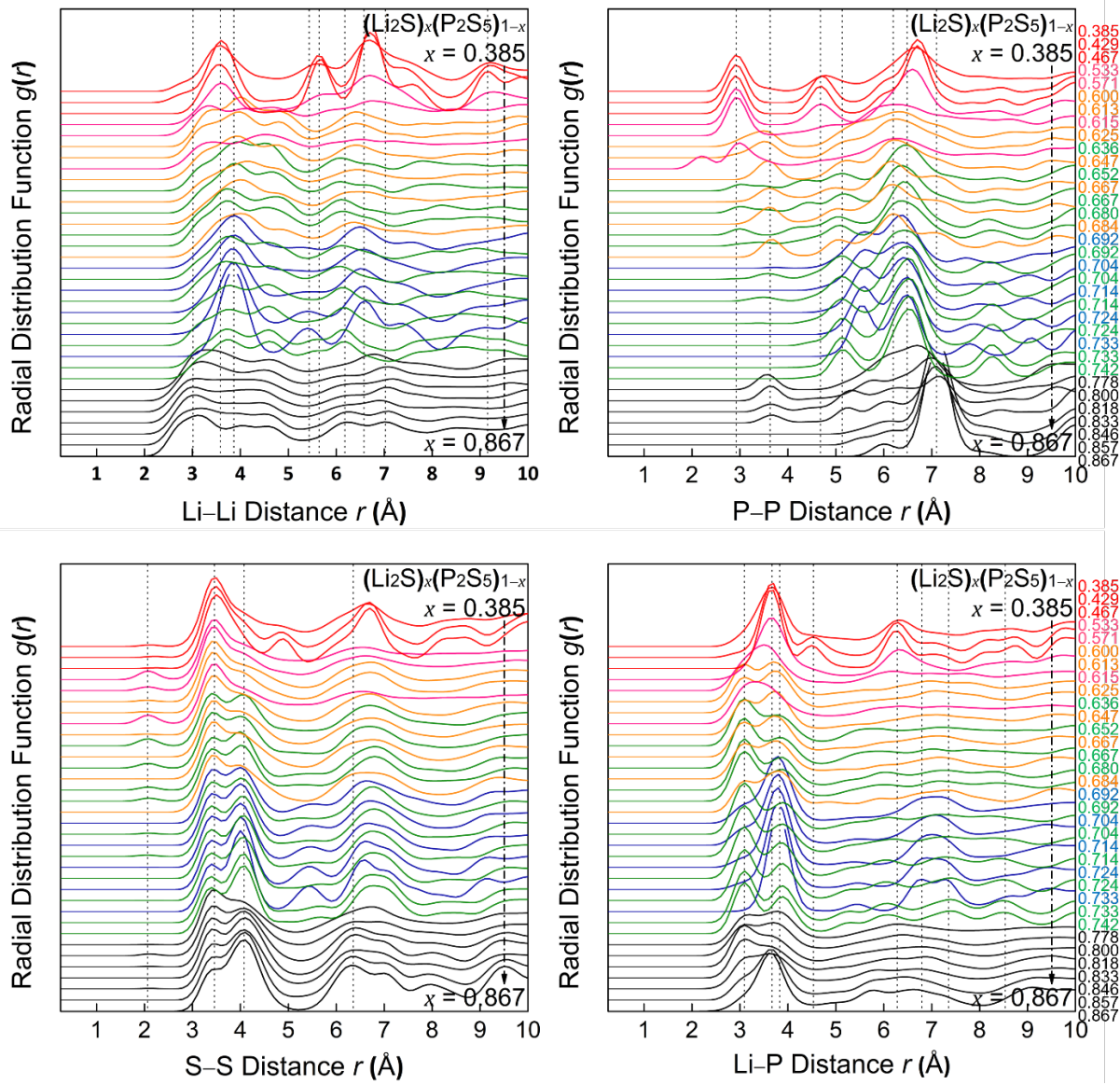

**Figure S1.** Calculated Li–Li, P–P, S–S and Li–P radial distribution functions (RDF) of glass-ceramic (*gc*) $(Li_2S)_x(P_2S_5)_{1-x}$ (*gc*-LPS) phases with changing compositions from x = 0.867 to x = 0.385. Each line is an average RDF of ten lowest energy structures at certain composition. Since the *gc*-LPS structures are generated with genetic-algorithm (details in Methods Section), the color represents their parent crystalline structure (*i.e.*, black: $Li_7PS_6$, blue: $\gamma$-$Li_3PS_4$, green: $\beta$-$Li_3PS_4$, orange: $Li_7P_3S_{11}$, pink and red: $LiPS_3$). The dashed lines represent measured RDF for crystalline phases.[1–8]

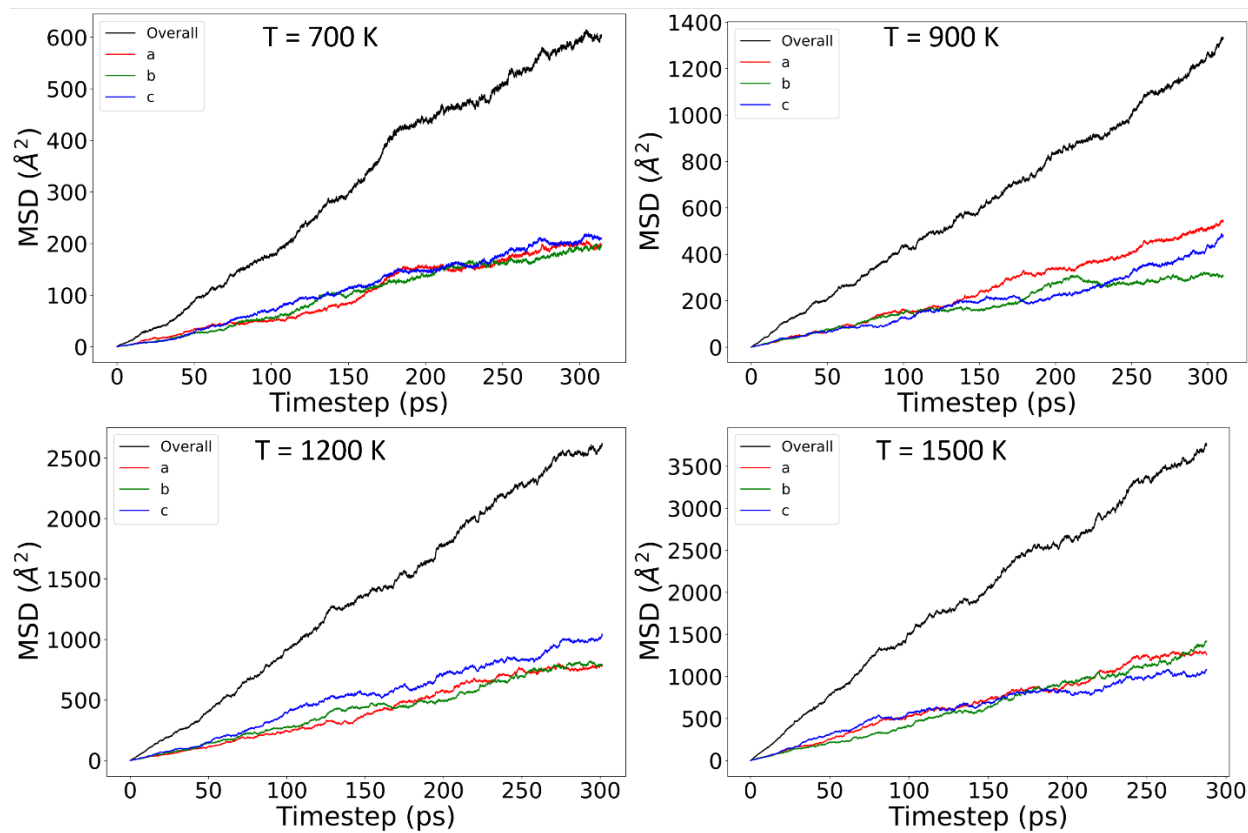

**Figure S2.** The mean-squared-displacement (MSD) of Li ions in *gc*-$Li_{42}P_{16}S_{61}$ at elevated temperatures (700 K, 900 K, 1200 K and 1500 K).